# Beyond Mobile Apps: a Survey of Technologies for Mental Well-being

Kieran Woodward, Eiman Kanjo, David J. Brown, T.M. McGinnity, Becky Inkster, Donald J. Macintyre & Athanasios Tsanas

**Abstract**— Mental health problems are on the rise globally and strain national health systems worldwide. Mental disorders are closely associated with fear of stigma, structural barriers such as financial burden, and lack of available services and resources which often prohibit the delivery of frequent clinical advice and monitoring. Technologies for mental well-being exhibit a range of attractive properties, which facilitate the delivery of state-of-the-art clinical monitoring. This review article provides an overview of traditional techniques followed by their technological alternatives, sensing devices, behaviour changing tools, and feedback interfaces. The challenges presented by these technologies are then discussed with data collection, privacy, and battery life being some of the key issues which need to be carefully considered for the successful deployment of mental health toolkits. Finally, the opportunities this growing research area presents are discussed including the use of portable tangible interfaces combining sensing and feedback technologies. Capitalising on the data these ubiquitous devices can record, state of the art machine learning algorithms can lead to the development of robust clinical decision support tools towards diagnosis and improvement of mental well-being delivery in real-time.

---

## 1 INTRODUCTION

Mental health problems constitute a global challenge that affects a large number of people of all ages and socioeconomic backgrounds. The World Health Organisation (WHO) [1] defines the well-being of an individual as being encompassed in the realisation of their abilities, coping with the normal stresses of life, productive work and contribution to their community. Hectic modern lifestyles contribute to daily stress and a general decline in mental health, as 59% of UK adults currently experience work-related stress [2]. This makes stress the leading cause of sickness-related absences from work, with about 70 million days lost each year at an estimated cost of £2.4 billion [2]. Furthermore, the Physiological Society [3] reported that 18-24 year-olds were the most stressed age group with students studying for higher degrees exhibiting considerable stress levels, where the majority (60.9%) of the high-risk undergraduate students rated their mental health as poor or very poor [4] showing the negative impact modern lifestyles are having on mental well-being.

Traditionally, clinical visits for physical and mental health assessment in chronic disorders are infrequent and intermittent, representing a very small time window into patients' lives, where clinicians are challenged to decipher the possible manifestation of symptoms and disease trajectory. Further problems are often encountered with patients' recall bias, when they are asked to retrospectively provide details and describe their symptoms. In many clinical fields patients are encouraged to use standardized clinical questionnaires, typically in the form of Patient Reported Outcome Measures (PROMs) or experience sampling [5] to understand the longitudinal variability of mental health symptom trajectory over months in-between clinical visits. A common problem encountered during clinical psychiatric assessments is that the questions asked about patients' mood, physical and mental health can be impacted by an unreliable autobiographical memory [6]. An alternative to traditional methods involves smartphone applications that can provide a variety of tasks including symptom assessment, talking therapies, psycho-education and monitoring the efficiency of treatment [7], [8].

Poor mental well-being often leads to physiological changes. For example, stress is defined as the non-specific response of the body to any demand for change, resulting in reduced heart rate variability [9], lower skin temperature [10] and increased skin conductance [11]. Technological advances have led to tangible interfaces in which a person interacts with digital information through the physical environment; these can incorporate sensors to measure physiological changes and help alleviate the stress people experience. This provides new opportunities to utilise non-invasive technology for behavioural health care in order to assess and aid mental health conditions such as anxiety and stress accurately in real-time. Multimodal interactions are currently used for a wide variety of purposes such as improving communication but mental well-being is an area where these interactions could have a profound impact [12], [13].

This study provides a literature survey and taxonomy that aims to explore the use of innovative interfaces that go beyond mobile applications to assess the potential of new

---

- *K. Woodward, E. Kanjo, D. Brown and T.M. McGinnity are with the Department of Science and Technology, Nottingham Trent University, Nottingham, UK. E-mail: {Kieran.woodward, eiman.kanjo, david.browm, martin.mcginnity}@ntu.ac.uk*
- *B.Inkster is with the Department of Psychiatry, University of Cambridge, UK. E-mail: becky@beckyinkster.com.*
- *D. Macintyre is with the Centre for Clinical Brain Sciences, Division of Psychiatry, University of Edinburgh, UK.*
- *A. Tsanas is with the Usher Institute, Edinburgh Medical School, University of Edinburgh, UK. E-mail: atsanas@ed.ac.uk.*

technologies and how they can be utilised to improve mental well-being. This survey explores all aspects of mental well-being recognition including *stress*, *depression* and *emotion recognition*. Emotion recognition differs from stress detection as it involves measuring the response to a particular stimulus (person, situation or event), usually intense, short experiences the person is aware of [14]. On the contrary, stress recognition involves detecting a reaction where individuals are subject to demands and pressures which do not correspond to their knowledge and abilities, challenging their handling capabilities [15].

Traditional methods to assess and improve mental well-being are first examined and then the technological alternatives are explored aiming to address the following research questions:
1. Can technology supplement traditional mental well-being assessment techniques?
2. Can machine learning be utilised to improve mental well-being classification?
3. How can behaviour changing tools be used to help improve mental well-being?
4. Can a combination of sensing and feedback technologies be used to improve mental well-being in real-time?

After these five highlighted areas have been reviewed, the challenges, tools, and opportunities modern technological advancements present for mental well-being are discussed.

## 2 A Taxonomy of Mental Well-Being Technologies Research

### 2.1 Traditional Assessment Tools and Techniques

Traditional methods used to assess mental well-being often utilise self-reporting for example, when people record their emotions and stresses in a diary that can be assessed and monitored to help establish stressful triggers [16], [17], or the use of validated questionnaires to measure daily life stresses, symptoms, etc. Examples of questionnaires include the Positive and Negative Affect Schedule (PANAS) [18], Quick Inventory of Depressive Symptomatology (QIDS) [19] and the Patient Health Questionnaire (PHQ-9) [20].

Diagnostic interviews are performed by psychiatrists/care professionals by asking service users and their friends or family about their symptoms, experiences, thoughts, feelings and the impact they are having. Diagnostic interviews allow for a diagnosis to be made according to standard classification systems such as ICD-10 [21] and DSM-5 [22] and these are used in conjunction with a biopsychosocial formulation to construct a management plan, which can include talking therapies which teach people to learn new behaviours, and develop greater resilience (e.g. to cope with stressful events) [23], [18]. Discussions with trained experts leads to potentially identifying underlying problems and can be used as treatment by teaching people new behaviours (e.g., to cope with stressful events).

Self-reporting diaries can take considerable time to assess as they must be completed over a long period to gain useful insights [24]. Also, symptom self-reporting can often be inaccurate due to poor recall; a study investigated how accurately individuals self-reported the number of fruit and vegetables eaten, with accuracies ranging from 40.4% to 58% [25]. Additionally, all of the traditional assessment methods require people to be aware of their mental health and actively seek help which often many forego due to fear of social stigma and lack of available resources [26], [27]. A technological alternative that could actively monitor patients' mental health state and provide methods to improve their mental well-being would be beneficial as it could improve accessibility to mental health tools [28].

### 2.2 Technological Supplements to Traditional Assessment Techniques

Can technology supplement traditional mental well-being assessment techniques?

#### 2.2.1 Overview of mHealth apps

With the high prevalence of smartphone ownership [29] access to treatment which is flexible and fits in with people's lifestyles is greatly enhanced [30]. Those at risk of mental health problems often have difficulty accessing quality mental health care [31] especially when symptoms first manifest [32] demonstrating the need for more accessible help. An Australian survey found that 76% of people would be interested in using mobile phone apps for mental health monitoring and self-management [33], illustrating the high demand for mHealth apps because of their convenience and accessibility.

Many apps have been developed to modernise and advance existing practices of recording mental well-being. Numerous mental health diary apps are available to download, although these are effectively digital representations of existing self-reporting diaries using new techniques such as the touchscreen, volume buttons and monitoring notifications [34], [35], [36]. However, using a phone in public is more socially acceptable than completing a paper form allowing monitoring to be completed discreetly in real-time, unlike paper forms which are often completed retrospectively, resulting in less accurate data being recorded [24]. A problem many apps face is the frequency for eliciting PROMs which may underrepresent the true symptom fluctuation. Given that mood is highly variable, clinically useful information is likely in the daily fluctuations of mood for many cohorts suffering from mental disorders. Previous research demonstrates the possibility of eliciting daily responses to assess mental health with very good adherence over a 1 year period [37] demonstrating the feasibility of longitudinal daily PROMs engagements by two cohorts diagnosed with bipolar disorders and borderline personality disorders. More recently, chatbot apps are being developed to assess mental well-being, in some cases by mimicking conversation with users via a chat interface [38] thus removing the requirement to continuously self-report. A survey conducted on 5,141 participants in the age range 16-24 years showed nearly two thirds would be comfortable with a chatbot giving them a diagnosis [39]. Chatbots can utilise artificial intelligence to reduce their reliance on predefined scripts and deliver individualised therapy suggestions based on linguistic analysis and enhance user engagement [40].

Furthermore, chatbots can generate emotional responses by using context sensitive advanced natural language-based computational models to detect user state and emotions and continuously provide personalised responses [41]. However, fully generative models for chatbots can result in hurtful comments on sensitive topics such as race [42] and mental health [43], [44], [45] which cannot be permitted in the domain of mental well-being as in this field, we must go beyond striving to pass the Turing test to additionally prioritise safety. It is of central importance that ethics and safety are constantly considered in this field, especially when working with young and vulnerable populations [46].

Text-based conversational chatbots can go beyond assessing mental well-being with some actively aiming to improve users' well-being. Wysa [47] and Woebot are two such chatbots that participants have found to be helpful and encouraging resulting in mood improvements [48]. Other mental well-being chatbots show positive reception of the intervention but also demonstrate the potential for artificial intelligence to understand the meaning of sentences without relying on pre-programmed keywords as this is a common criticism of chatbots [49]. There is increasing interest in this type of bot-based interactive support as Wysa has been downloaded over 500,000 on the Google Play store alone [50]. iOS and Android app stores allow any developer to publish mental health apps without any precautionary checks or safeguards that go beyond standard malicious program assessment, such as also verifying whether apps have been scientifically evaluated.

Figure 1 provides a summary of the five top rated popular mental health apps (in the UK app store as of January 2019), each of which has an overall rating of at least 4.4 out of 5. For comparison and benchmarking we also present *Wellmind*, an app developed by the National Health Service (NHS) in the UK. The six apps have been developed by a wide range of organisations with varying levels of features and effectiveness. Although many of these apps such as 'Calm' and 'What's up?' have engaging interfaces and are fairly intuitive to use, we stress that typically there is no scientific evaluation to confirm their effectiveness. App stores could be more rigorous in their testing and approval of mental well-being apps to prevent erroneous conclusions being drawn by individuals, which could potentially lead to detrimental impact on people's mental health. We envisage this may be an area where new developments might require health apps to indicate whether they have been externally certified as fit-for-purpose to make users aware.

Mental health apps are also increasingly becoming profitable businesses. For example, Calm, a meditation app which is free to download and use has recently been valued at $1 billion [51] even though there have been no clinical trials or evaluation to confirm the mental well-being benefits of using the app. More worryingly, Apple and Google have endorsed Calm by making it the 2017 app of the year and the 2018 editor's choice respectively [52], which could create a strong impetus towards people adopting the app despite the lack of scientific evidence supporting its use. There are studies demonstrating the benefits of mindfulness technology interventions [53], [54] but hitherto no evaluation has proved the benefits of Calm over evaluated competing apps (some of which have been scientifically validated).

Similarly, Calm Harm, an app designed to prevent self-harm, is featured on the NHS digital library [55] and while the app has been developed by a psychologist using principles of practice there have been no clinical trials or evaluation to confirm efficacy. The presence of Calm Harm on the NHS digital library suggests this is a legitimate, evidence based app. The NHS digital library categorizes apps using three distinct badges: (i) approved, (ii) being tested and (iii) no badge [55]. Calm Harm has received no badge indicating it meets NHS quality standards for safety, usability or accessibility and it is not currently being tested by the NHS for clinical effectiveness. The badge system used by the NHS allows any app meeting their unpublished standards to be prominently displayed and easily misrepresented as clinically tested.

Headspace currently has over ten million downloads on the Android Play store alone, underlining the immense popularity of mobile well-being apps. Unlike Calm, Headspace has published research findings demonstrating it can help reduce stress by 14% [56], increase compassion by 23% [57], reduce aggression by 57% [58] and improve focus by 14% [59]. However, most of these studies were small scale with the longest period people were followed being just thirty days. Another research study reported that using the app over a six week period resulted in no improvements in critical thinking performance [60]. Additionally, there has been no follow-up after the initial studies and as some studies lasted as little as ten days it raises some concerns as to whether the positive outcomes from the app may only be apparent during an individual's initial period of use.

| | | | | |
|---|---|---|---|---|
| 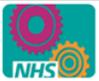 | **Wellmind - NHS** 3.4*** Record Feelings, Advice and Relaxing audio Developed by reputable organisation but offers little functionality other than the ability to read general information and record limited moods. | | 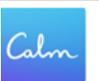 | **Calm - Calm.com** 4.6***** Range of activities to help comfort, distract, release, breathe and more. The app provides a variety of tasks to complete all within different categories but these tasks have not been tested to ensure effectiveness. |
| 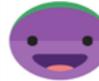 | **Dayio - Relaxios.r.o** 4.8***** Simple app that provides an effective way to monitor moods and what might impact mood over time, much like traditional self-reporting but easier to access. The ability to customise the moods is useful and a feature many other apps do not offer. | | 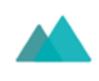 | **MoodPath - MoodPath UG** 4.6***** Tracks mood, offers mental health assessment and information on detection and treatment. Has very limited functionality. It is intuitive through the use of large simple icons and provides a mental health assessment after 14 days. The app also provides potentially useful statistics about mood over time. |
| 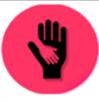 | **Whats Up? – Jackson Tempra** 4.4**** The app has a large number of features but is very unintuitive with a complex user interface relying on custom icons. There is little research about how well the included help such as breathing control, grounding and uplifting quotes work. | | 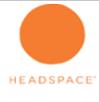 | **Headspace - Headspace** 4.6 ***** Provides guided meditation to help reduce stress and anxiety and improve focus and sleep. The app has a wide range of guided mediation available with useful goals and statistics to make monitoring progress easy. However there is little evaluation to prove its effectiveness. |

Fig. 1: Summary of indicative popular mental health apps in the Google Play store [48] compared with Wellmind, an app developed by the NHS.

Figure 2 presents the total number of global downloads and average rating of the six most downloaded mental health apps on the Google play platform. The total number of downloads varies widely as 'Headspace', 'Calm' and 'Daylio' make up the vast majority of downloads with a combined total of 25 million whereas next most popular apps only amass 500,000 downloads each, showing that receiving favourable reviews does not necessarily lead to mass downloads. Evaluated apps developed by respected organisations also do not necessarily result in popularity as 'Wellmind' developed by the NHS has only been downloaded around 10,000 times and received an average rating of 3.4 out of 5, reflecting users' preference of usability and functionality.

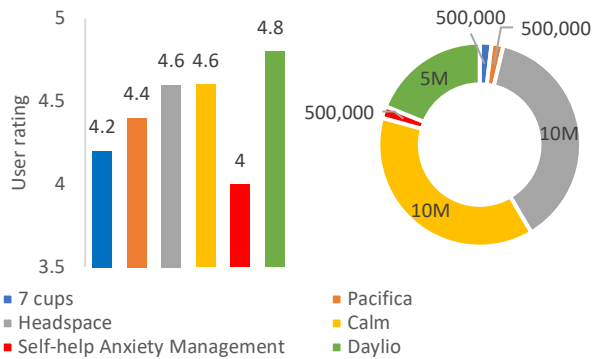

Fig. 2: Comparison of average rating (left) and total global downloads (right) of the six most downloaded mental health apps from the Google Play store.

Both the iOS app store and Google play do not have a dedicated category for mental health apps meaning they are combined with other health & fitness apps increasing the difficulty of finding relevant apps. Figure 3 shows the subcategories of the top 100 free and paid for health and fitness apps on the UK iOS App store in September 2018. The majority of apps within the health & fitness category are dedicated to exercising with only a small proportion of apps for stress or mood monitoring and these apps were generally lower in the charts obscuring them from users. App stores could improve the visibility of tested mental health apps through a dedicated mental health category which may facilitate the uptake of well-established apps which have received positive feedback from users.

Additional apps have been developed by researchers that actively aim to improve mental health and well-being such as mobile stress management apps that use stress inoculation training to prepare people to better handle stressful events. Studies show stress inoculation apps were consistently successful in reducing stress in participants and increasing their active coping skills [61], [62]. Grassi et al. [63] demonstrated that mHealth apps are not only capable of augmenting traditional techniques to help monitor conditions but they can also be used to educate users on techniques to actively improve their mental well-being.

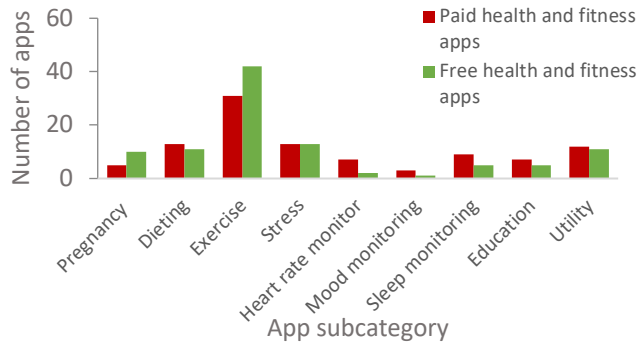

Fig. 3: Categories of the top 100 health and fitness apps in the UK iOS app store.

A smartphone app, FOCUS, has been developed to proactively ask users with schizophrenia about their mood, feelings and well-being multiple times each day to provide relevant coping strategies [64]. This allows the app to go beyond traditional self-reporting as it educates users on methods to help immediately after an issue has been reported which is only possible using technology that people have continuous access to such as smartphones. FOCUS demonstrated a reduction of positive symptoms of schizophrenia and depression, when trialled by 33 participants over 4 weeks. A common issue with mental well-being apps is low user engagement. However, FOCUS was used by participants on 86.5% of days, averaging 5.2 times each day over 30 days and Oiva, a mental well-being training app [65] was on average used every third day for 12 minutes over a 30 day period demonstrating the possibility for mental well-being technologies to be highly engaging.

While apps could be considered as an alternative to seeking professional help some apps have been designed to work in conjunction with clinicians such as Post-Traumatic Stress Disorder (PTSD) coach. The app allows users to learn more about PTSD, track symptoms, set up a support network and provides strategies for coping with overwhelming emotions. 10 US veterans with PTSD were assigned to use PTSD Coach independently while another 10 used the app with the support of their primary-care providers [30]. At the end of the trial, seven of the ten patients using the app with support showed a reduction in PTSD symptoms, compared with just three of the patients who used the app independently. Apps used with care providers show more potential for effective treatment in the small sample trials although this still requires users to actively seek help [66].

Pairing apps with psychiatrists' and psychologists' support has been shown to be successful resulting in a range of apps using content explicitly created by psychiatrists such as Rizvi et al. [67] who developed the app DBT Field

Coach to provide instructions, exercises, reminders, games, videos and messages to help people cope with emotional crises. The results of that study demonstrate the 22 participants used the app frequently over at least 10 days and it was successful in reducing intense emotions, reducing substance use cravings and improving symptoms of depression without the need to visit a clinician [67]. This app again shows the success of apps utilising psychiatrists and clinicians although as this app only used content created by psychiatrists, it negates the need to visit clinicians increasing accessibility. Mobile health apps provide many advantages over traditional techniques including improved accessibility, real-time symptom monitoring, reduced cost and reduced barriers to access [68]. One of the main shortcomings of available smartphone apps is the lack of personalised features as many treatments and strategies have to be individually tailored [69].

*2.2.2 Tangible Interfaces*

An alternative method to enhance existing techniques is through the use of tangible interfaces which are user interfaces in which a person interacts with digital information through the physical environment. This presents new opportunities as Matthews and Doherty [70] and Niemantsverdriet and Versteeg [71] have reported that people are more likely to create stronger emotional attachments with physical devices rather than digital interfaces such as apps.

These tangible devices provide a technological alternative to traditional self-reporting, allowing users to report their current mental well-being in real-time. Emoball [72] is one such device that allows users to record their mood by squeezing an electronic ball making them conscious of their current mood. While this device only allows users to report a limited number of emotions participants did believe mental well-being and education were the areas where devices to report emotions could be of most use. A smaller, portable device that works similarly is Keppi [73], which allows users to squeeze to record low, medium or high pain.

Another tangible approach to self-report is the mood TUI [74], which as well as allowing users to record their emotions also collected relevant data from the user's smartphone including location data and physiological data such as heart rate. Participants found the use of a tangible interface very exciting, although when the device was tested with users, they felt it was too large and they would lose motivation to continue using it for an extended period. This feedback shows the use of tangible user interfaces excites users, but the design and functionality must be prioritised. Mood sprite [75] is another handheld device developed to help people suffering from anxiety and stress by using coloured lights and an infinity mirror to assist with relaxation. The device records the time users create new sprites allowing them to be revisited much like a diary again showing ways in which tangible interfaces can accompany traditional techniques to make treatment more accessible and user centric. The device educates users similarly to traditional self-reporting diaries by allowing them to recall their emotions but is more engaging with different coloured lights representing different times and moods promoting continued use. However, a common issue with mental health tangible interfaces is that they remain largely unproven and even those that have been trialled with users such as Mood sprite have been done so in small-scale trials that lack statistical power.

Subtle Stone [76] is a tangible device that allows users to express their current emotion through a unique colour displayed on a stone, limiting the number of people with whom users share their emotions with. Subtle Stone was tested with eight high school students in their language class with the teacher able to view the data in real-time using an app. The study showed the use of colours to represent emotions was well received with students liking the anonymity it provided along with finding it easier to use than words. Subtle Stone both allows users to communicate their emotions privately and monitor their own emotions over time demonstrating clear advantages over traditional self-reporting methods.

A tangible interface used to detect stress in real-time without the need to self-report is Grasp, which was tested with anxious participants in a dentist's office [77]. Participants were able to squeeze Grasp whenever they felt stressed and the device detected how much pressure was exerted and displayed this data on a mobile app. Force sensors have also been used to create a tactile ball that allows for the manipulation of music by squeezing different areas of the ball along with movement detected by an accelerometer [78]. The research concluded squeeze music could successfully be used for music therapy with children as it promoted positive emotions through tactile input and music. Sensors such as force sensors have been shown to provide an intuitive method of interaction for tangible user interfaces and show the possibility for additional sensors to be utilised when educating, detecting and improving mental well-being that is not possible when using smartphones or traditional techniques.

*2.2.3 Evaluation of Discussed Technologies*

The rise and popularity of mental well-being smartphone apps highlights their potential usefulness. However, we stress that many existing mHealth apps have not been tested in scientifically rigorous research studies despite the fact that many have millions of users. Mobile apps are likely most beneficial when used to display clinically approved content or replace traditional techniques such as self-reporting (paper-based) diaries with technological alternatives. However, caution should be exercised when developing apps that aim to improve mental well-being without being first thoroughly tested.

There are multiple tangible interfaces that go beyond apps by utilising various sensors to provide a variety of purposes including self-reporting of emotions, relaxation and communication. When developing tangible mental well-being interfaces, the design needs to be carefully considered to ensure it is effective and not damaging. Guidelines [79] have been introduced to ensure mental health technologies are successfully developed. The guidelines address the design process, the development of the devices and evaluation procedures. The guidelines include designing for outcomes with health care professionals, making the system adaptable and sustainable, and also providing flexibility in the delivery of support. The guidelines are relevant to a wide range of mental well-being technologies from monitoring devices to biofeedback devices.

mHealth apps have multiple benefits due to their constant accessibility, while tangible interfaces provide new, intuitive ways to interact and visualize data. Overall tangible interfaces and apps provide new opportunities to enhance existing assessment methods as the convenience and additional functionality lead these technological alternatives to improve the reporting and communicating of mental well-being.

## 2.3 Sensing Mental Well-Being

Can machine learning be utilised to improve mental well-being classification?

Advances in deep learning have resulted in benefits far beyond those of machine learning, including the capability to classify the raw sensory data overcoming the laborious process of manual feature engineering and presenting the extracted features to a statistical learner.

There are two main neural network types: Convolutional Neural Networks (CNNs) and Recurrent Neural Networks (RNNs). The main difference between CNN and RNN is the ability to process temporal information. They are structurally different and are used fundamentally for different purposes. CNNs have convolutional layers to transform data, whilst RNNs essentially reuse activation functions from other data points.

RNNs relying on Long Short-Term Memory (LSTM) are especially valuable for use with sensor data as they are fundamental in distinguishing similar data which differ only by the ordering of the samples which can often dictate differences in mental health [80].

CNNs have traditionally been used to classify images and speech due to their ability to scale invariance of a signal, however recently their application has been expanded to classify raw sensor data [81], [82]. The inputs in a convolutional layer connect to the subregions of the layers instead of being fully connected as in traditional neural networks. As the inputs of a CNN share the same weights, they produce spatially correlated outputs.

Deep learning advances create the potential to improve the performance of mental well-being classification. The following sections explore the classification of mental well-being using data collected from mobile applications, multimodal physiological sensors, text, speech, images and video.

### 2.3.1 Mobile App Approaches

Apps have been shown to enhance traditional PROMS-based assessment techniques and by utilising sensors within phones, the capability of apps is further enhanced: the apps may potentially provide a more holistic picture using passively collected data. Smartphones are capable of collecting a vast amount of data such as location, motion and phone use which can result in many features being extracted to train machine learning algorithms. It is possible to use the data collected from smartphones to determine emotions with a 70% accuracy utilising machine learning to process the data [83]. Automatically inferring emotions based on smartphone use is extremely valuable in determining mental well-being and can provide new clinical insights from passively monitoring users' behaviour.

In addition to using a phone's sensors to detect mental well-being, it may be possible to use a phone's touchscreen to sense stress. Using an infrared touchscreen to measure photoplethysmograph (PPG) it was possible to recognise stress with accuracies of 87% and 96% across two tests, a vast improvement upon previous touchscreen based stress detection [84]. However, infrared touchscreens are rarely used especially within smartphones, although the possibility of measuring stress through capacitive touchscreens could have significant impact.

Smartphone apps have also been paired with wrist-worn sensors to infer mental well-being by allowing for a high magnitude of data to be collected [85]. The collected data was expressed using 15 multimodal features ranging from physiological data such as skin conductance to phone usage data such as screen time duration. The 15 sets of features were then trained with a variety of classifiers and the accuracy of the different features were examined for each classifier. The system was capable of detecting stress with a 75% accuracy, with some of the features such as increased acceleration during sleep and high evening phone use being more beneficial than others in determining stress. Similarly, a wrist sensor along with a mobile app and a self-reported PHQ-8 and PHQ-4 depression scores were used to quantify depression symptoms in 83 undergraduate college students across two 9-week periods by measuring phone use, heart rate, sleep and location [86]. The study concluded students who reported they were depressed were more likely to use their phone at study locations, have irregular sleep, spend more time being stationary and visit fewer places. They demonstrated that they could automatically detect depression with a 69.1% precision when evaluated against the PHQ-4 depression subscale [87] and that this could be improved if additional physiological sensors were included. In addition to physiological sensors, location could be used to assess mental well-being as movement patterns and uncertainty in visits has been shown to be predictive of the outcomes of the Quick Inventory of Depressive Symptomatology (QIDS) [88]. These studies demonstrate the potentially powerful combination machine learning, sensors and mobile apps provide when tested in high quality trials to automatically determine stress levels.

BreathWell [89], which has been developed for Android Wear smartwatches has been designed to assist users in practising deep breathing to reduce stress from PTSD although the app has limited functionality to determine stress as it only uses the wearer's heart rate. Despite the limited functionality, all seven participants believed the app could help them and preferred the app being incorporated into a wearable device making it more convenient to use although the extent of the trial was extremely limited.

Figure 4 shows widely used contemporary sensors contained within smartphones and smartwatches and how

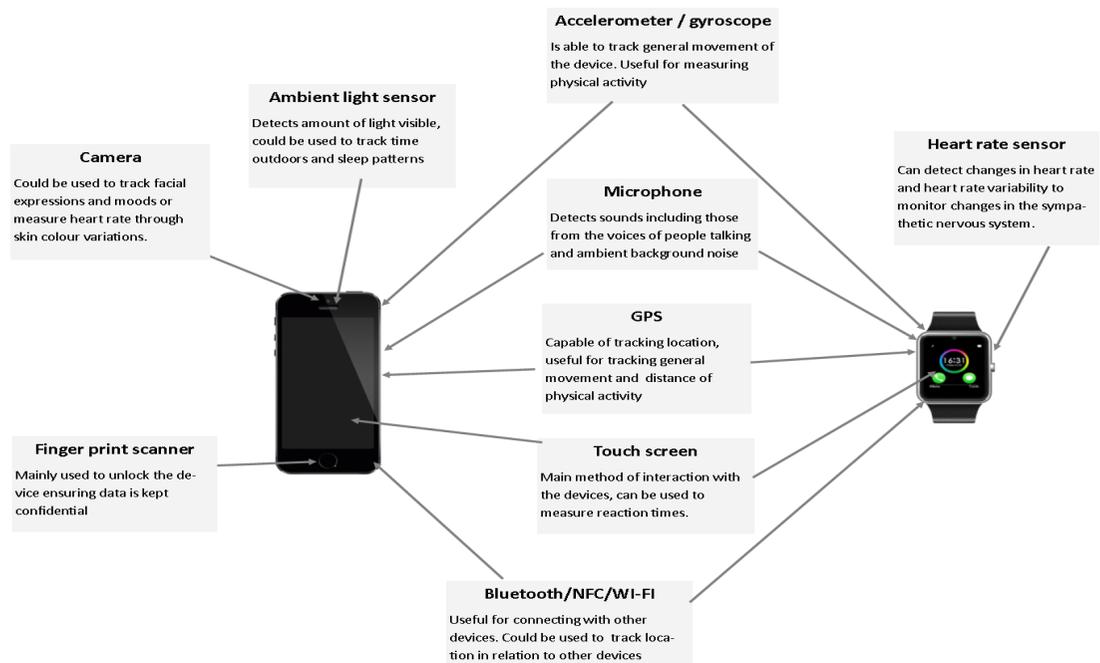

Fig. 4: Possible uses of smartphone and smartwatch sensors in relation to mental well-being apps could further capitalize on the data collected to assess mental well-being more accurately. Some sensors are already widely utilised such as heart rate as this can be directly associated with mental state, but other commonplace sensors such as the camera, GPS, and accelerometer could be used more effectively within mHealth apps.

### 2.3.2 Multi-Modal Physiological Sensor Approaches

Machine learning is vital to accurately infer mental well-being. There are numerous sensors that when combined with sufficiently trained machine learning classifiers can be used to assess mental well-being in real-time.

Non-invasive physiological sensors present the most significant opportunity to assess mental well-being. The main measures for stress are brain wave activity, Galvanic Skin Response (GSR) and Heart Rate Variability (HRV) [90]. GSR is often used to detect mental well-being as it directly correlates to the sympathetic nervous system [91]. A CNN has been trained to classify four emotions, relaxation, anxiety, excitement and fun using GSR and blood volume pulse signals [81]. The deep learning model outperformed standard feature extraction across all emotions achieving accuracies between 70-75% when the features were fused.

Near-Infrared Spectroscopy is a non-invasive sensor that measures oxyhaemoglobin and deoxyhaemoglobin, and research has shown this can be used to detect mental stress similar to GSR [92] but is more challenging to use outside of laboratories due to its large size and placement on the forehead. Stress can also be detected from brain activity using ElectroEncephaloGrams [93] (EEG) as Khosrowabadi et al. demonstrates using eight channels to classify students' stress during exams with over 90% accuracy [82]. A CNN with channel selection strategy, where the channels with the strongest correlations are used to generate the training set, has also been used to infer emotion from EEG signals [94]. The model achieved 87.27% accuracy, nearly 20% greater than a comparative model without channel selection strategy. Similarly raw EEG signals have been used to train a LSTM network achieving 85.45% in valence [95].

A wearable device that aimed to detect stress measured ElectroCardioGram (ECG), GSR and ElectroMyoGraphy (EMG) of the trapezius muscles [96]. Principal component analysis reduced 9 features from the sensor data to 7 principal components. 18 participants completed three different stressors; a calculation task, a puzzle and a memory task with a perceived stress scale questionnaire completed before and after each task. The principal components and different classifiers were used to detect stressed and non-stressed states with an average of almost 80% classification accuracy across the three tests compared with the questionnaire results. However, this study only detected two states; stressed and non-stressed and was conducted in a controlled environment so it is not known how accurate it is in real-world setting as physiological signals can be affected by factors other than mental well-being.

Furthermore, LSTM networks have been used to classify other objective data including GSR, skin temperature, accelerometer and phone usage data to infer stress. The LSTM model achieved 81.4% accuracy, and outperformed the other Support Vector Machine (SVM) and logistic regression models [97]. LSTM networks have been used to classify EEG signals inferring emotions with 81.1% accuracy when using the context correlations of the feature sequences [98]. A CNN and LSTM have been combined to allow raw data to be classified more accurately [99], [100]. This deep learning approach is capable of using raw data to automate the feature extraction and selection stages. This approach to classifying emotions from physiological, environmental and location data outperformed traditional multilayer perceptrons by over 20%. The ad-hoc feature extraction by the CNN matched or outperformed models with the features already extracted showing the clear advantages of using deep learning approaches.

HRV is commonly used to assess stress as this is the variation in time between heartbeats meaning the lower the HRV, the more likely the user is to be stressed [101]. It is possible to measure HRV using electrocardiograms [102] but in 1997 it was found that finger pulse amplitude decreased significantly during mental tasks [103] leading to HRV being accurately measured using PhotoPlethysmoGraphy (PPG) which is easier and more cost-effective to use than ECGs as it only requires one contact point. There are three types of PPG; transmitted, reflected, and remote. Transmitted signals are often used in medical monitoring [104], whilst remote signals use cameras to detect changes to measure HRV by monitoring skin colour changes [105], [106]. Reflected measures the signal reflected from a LED using light sensing photodiodes to measure HRV, making this the smallest and most convenient method to use in tangible interfaces [107].

Both GSR and HRV were used in a wearable device to measure stress during driving [108]. The wearable device took measurements over a 5-minute period to detect stress levels with an accuracy of 97.4% and found that HRV and skin conductance are highly relatable making them extremely useful in detecting mental state. The ability to use sensors to measure HRV and skin conductance allows for small wearable devices to accurately determine stress levels in real-time and should be further utilised to detect stress, anxiety and mental well-being. However, physiological signals do not account for the context in which the devices are used as the context can play a significant role in the users' perceived stress levels meaning additional environmental sensors may also be required [109].

Another non-invasive sensor that has previously been used to detect stress is skin temperature as it can indicate acute stressor intensity [110]. One study [111] used a wearable device that contained multiple sensors including skin conductance, skin temperature and motion and provided it to 6 people with dementia and 30 staff in a nursing home for 2 months. The device aimed to automatically detect stress and categorise it into one of five levels, the accuracy for each of these levels varied from 9.9% to 89.4% showing an extremely wide variation. This was due to the threshold setting: when it was raised, fewer events were classified as stress because of the more challenging criteria, in turn, increasing precision. Accurately assessing stress levels is extremely useful as it allows for only the required stress to be recorded depending on whether all data or a higher accuracy is required.

While tangible interfaces paired with machine learning have shown the ability to infer mental well-being state in limited trials, the new computational advancements discussed have demonstrated high accuracy when classifying data and can be successfully ran from wearables and smartphones providing opportunities to more accurately detect mental well-being in real-time. Combining all these data streams along with intelligent algorithms may greatly advance the field of digital psychiatry and mental health.

### 2.3.3 Text, Speech, Images and Video Approaches

Recent studies have demonstrated that mental well-being can be assessed through physiological sensors and there is increasing evidence that well-being can be assessed through mining text using natural language processing. For example, we could mine text that comes in the form of social media posts. When detecting depression on Reddit an accuracy level of 98% was achieved when vector-space word embeddings were combined with lexicon based features [112]. Depression on Twitter has also been explored achieving 81% accuracy when using a bag of words approach where the frequency of each word is counted using a dataset of 2.5 million tweets crowdsourced over one year [113]. Twitter data has also been used to infer PTSD, depression, bipolar and seasonal affective disorder and when tested a log linear model was successfully able to separate the control data from diagnosed data for each disorder [114]. Similarly, Facebook posts can be mined to predict depression. By comparing Facebook posts with medical reports from 683 patients it was possible to predict depression with 69% accuracy [115]. Blog posts have been used to train classifiers to infer six different emotions with 84% accuracy [116], while an SVM classifier achieved 69% accuracy when classifying emotions from messages [117]. Emojis from Twitter have also been used to infer emotion using an SVM classifier although final f1 scores were between 10%-64% for the 6 emotions [118]. A gated RNN has similarly been used to classify 24 emotions with 87.58% accuracy from tweets using hashtags as emotion labels, which increased to 95.68% when classifying 8 primary emotions [119]. Stress and anxiety have also been inferred through text. A hybrid multi-task model improved stress classification from social media posts by 10% [120]. Similarly, correlations between social media posts and stress concluded domain-adapted features outperformed sociodemographic features traditionally used in machine learning models [121]. A lexical approach and a set of rules have also been used to infer stress from tweets proving a more practical application although less accurate than machine learning models [122].

Recent advancements in artificial intelligence have also enabled mental well-being to be inferred from speech signals. A three minute speech test has been used to identify children with anxiety and depression [123]. By using a speech test that is simple for children to complete and logistic regression and SVM models it was possible to detect anxiety and depression with 80% accuracy compared with self- and parent-reported questionnaires and diagnostic interviews. The majority of the previous work utilising speech to sense well-being uses speech collected in controlled environments. However, datasets containing speech of acted emotions and authentic emotions from television talk shows have been used with an estimator to define emotions on their valence, activation and dominance [124]. A k-nearest neighbour classifier was used to classify emotions with up to 83.5% accuracy. Additionally, hidden markov models have been used to infer six emotions from the speech of 12 speakers achieving an average accuracy of 78% [125]. Stress can also be inferred from speech as a LSTM classifier trained with data from 25 participants achieved an average accuracy of 64.4% [126].

Speech has also been considered for the long term monitoring of people with a bipolar disorder [127]. Long term monitoring involved the continuous collection of labelled structured speech and additional unstructured speech via phone calls. 24 features were extracted from the data and used to train an SVM with linear and radial-basis function kernels. The classifier achieved accuracies of 81% for hypomania and 67% for depression using the labelled dataset however when tested on the unstructured dataset accuracies reduced to 61% and 49% for hypomania and depression respectively. This demonstrates the difficulty of classifying ecologically valid long-term speech compared with sensors, which are substantially simpler to use in-situ.

In addition to speech and text it is also becoming increasingly popular to infer mental well-being from video and images. Facial actions have been used to detect depression in 57 participants using manual Facial Action Coding System (FACS) and active appearance modelling (AAM) [128]. A SVM classifier was used to detect depression with 88% accuracy for FACS and 79% for AAM compared with clinical diagnosis. A SoftMax regression-based deep sparse autoencoder network has been used to infer 7 emotions achieving up to 89.12% accuracy, a 13.37% improvement over a traditional SoftMax regression classifier [129]. Transfer learning has been used to improve facial emotion recognition within small datasets improving accuracy by 16.47% [130]. Similarly, a Raspberry Pi has been used to enable the real-time classification of five emotions from images, achieving 94% accuracy [132] and CycleGAN used a generative adversarial network to improve the performance of facial emotion recognition from an unbalanced dataset by up to 10% [131]. Furthermore, depression and anxiety have been classified from social media profile pictures using multi-task learning [133]. Instagram photos were used to measure depression achieving an F1 score of 0.647, outperforming general practitioners average diagnostic success rate [134]. Instagram photos have also been used to uncover visual attributes of photos relating to mental health conditions including bi-polar, anxiety and depression related conditions [135].

Alternatively, video can be used to replace physiological sensors. By using video feeds of people's faces it is possible to measure heart rate and with the use of machine learning the error rate was reduced to only 3.64 beats/min demonstrating a potential alternative to the use of sensors [136]. Thermal imaging cameras have also been used to detect breathing patterns to infer stress using a CNN achieving 85.6% accuracy [137]. Furthermore, 3-D facial expressions and speech have been used to measure depression compared with the patient health questionnaire [138]. An LSTM classifier achieved 74.2% accuracy while a casual CNN achieved 83.3% accuracy showing its increased performance on long sequences.

The use of video and images to infer mental well-being demonstrates a high level of accuracy, but requires the use of multiple cameras to continuously record participants and hence is not currently suitable for real world environments. Speech shows greater potential for real world applications as it can utilise the microphone embedded within smartphones although it remains challenging to continuously record speech especially in noisy environments. The classification of text to infer mental well-being is both accurate and easy to complete as text messages and social media posts can be used to infer well-being in real-time.

### 2.3.4 Data Analytics and Datasets

Mental well-being inference relies on the collection of multi-modal data that holds information on individuals' mental states.

While machine and deep learning advances mental well-being inference it requires a large labelled dataset to initially train the models which can be challenging to obtain. Crowdsourcing [139] is often used to label images, video and audio data which can result in incorrectly labelled data used to train the models. Furthermore, even if the data is labelled by experts it might not always reflect the true internal state of the user. A hybrid approach of self-reporting and continuous data collection would enable more accurately labelled data to be collected but this relies on users continuously reporting their well-being [140].

Before data analytics can be conducted or machine learning models trained, a large labelled dataset is first required. The use of reliable datasets is necessary as models may demonstrate high performance during training but perform poorly when tested in the real-world. There are several published affective datasets containing a variety of data sources as shown in Table 1 below.

TABLE 1
AVAILABLE AFFECTIVE DATASETS

|  | Data Source | Users | Measurement |
|---|---|---|---|
| **Deap** [141] | EEG | 32 | Arousal, valence, like/dislike, dominance & familiarity |
| **AMIGOS** [142] | EEG, ECG, GSR | 40 | valence, arousal, familiarity, like/dislike, and emotions |
| **SEED** [143] | EEG | 15 | Emotion and vigilance |
| **CASE** [144] | ECG, BVP, EMG, GSR | 30 | Self-report valence and arousal |
| **SWELL-KW** [145] | HRV, GSR, body posture, facial expression, computer interaction | 25 | Task load, mental effort, emotion and perceived stress |
| **WESAD** [146] | HR, ECG, GSR, EEG, respiration, body temperature, & acceleration | 15 | Neutral, stress, amusement |
| **HCI Tagging** [147] | ECG, EEG, respiration amplitude, skin temperature, eye gaze, video, audio | 30 | Valence and arousal |
| **EmoBank** [148] | 10k words | N/A | Valence-Arousal-Dominance |
| **Sentiment140** [149] | 1.6m tweets | N/A | 4 Affective states |
| **CelebA** [150] | 202599 facial images |  | 40 Attributes |
| **BU-3DFE** [151] | 2500 3d facial expressions | 100 | 7 Expressions |
| **TESS** [152] | Audio of 200 target words | 2 | 7 Emotions |
| **RAVDESS** [153] | Audio & visual speech & song | 24 | 7 Expressions |

Signal processing can be used on large scale multi-modal datasets to identify hidden attributes from the raw sensor data. Signal processing techniques can be beneficial once raw sensor data has been collected as they have previously measured atypical speech for people with autism [154], measured depression using heartbeat dynamics [155] and detected common physiological signals associated with bipolar disorder [156].

Signal processing mobile frameworks simplify the process of analysing real-time signals. Frameworks been developed that aim to ease the collection of sensor data and ease the labelling of the data, that is required before data can be classified [157]. Another mobile framework augments social interactions by analysing smartphone sensor data in real-time to then provide live feedback improving users' behaviour [158]. Similarly, MediaPipe [159] is a framework that aims to assist the selection and development of multi-modal machine pipelines that has frequently been used for object detection. Signal processing mobile frameworks can be used to analyse physiological data [160], [161] greatly assisting the collection and processing of labelled multi modal data for mental well-being detection.

Table 2 below summarises all of the discussed approaches to infer well-being categorised by modality.

TABLE 2
MODALITIES FOR MENTAL WELL-BEING INFERENCE

|  | Depression | Stress & anxiety | Emotion | Bi-polar |
|---|---|---|---|---|
| EEG |  | [82] | [94],[95] |  |
| ECG |  | [96] | [98] |  |
| GSR |  | [96], [108], [111] | [81] |  |
| HR | [155] | [84] |  |  |
| HRV |  | [108] |  |  |
| Skin temperature |  | [111] |  |  |
| Smartphone usage |  |  | [83] |  |
| Smartphone & physiological | [86] | [97] | [85],[99], [100] | [156] |
| Text | [112],[113], [114],[115], [162] | [120], [121], [122] | [118],[116][119],[117] | [114] |
| Speech |  | [123], [126] | [124], [125] | [127] |
| Images and video | [128],[138], [135],[133], [134] | [133], [135], [137] | [129],[130],[132],[131] | [135] |

## 2.4 Technological Interventions

### 2.4.1 Virtual and Augmented Reality

How can behaviour changing tools be used to help improve mental well-being?

Numerous studies have shown Virtual Reality (VR) to help improve many psychological disorders including PTSD and anxiety by allowing patients to be exposed to stressful or feared situations in a safe environment [163], [164]. When using VR people are aware the situation is artificial allowing them to temporarily suspend their disbelief and be more confident in trying different approaches.

A pilot study at the University of Oxford demonstrated that virtual reality tools might reduce the delusional beliefs that are associated with schizophrenia and severe paranoia [165]. Participants experienced a lift or train simulation. The group that dropped their defence behaviours showed substantial reductions in their paranoid delusions, with over 50% no longer having severe paranoia within the simulated situation. Furthermore, a 19.6% reduction in distress in real-world situations was achieved. VR allows people to learn new approaches, helping improve their mental well-being in real-world situations although further research is needed to see if the benefits are maintained for more than the specific scenarios trialled [166].

Augmented reality (AR) has the capability to assist people in the real world by overlaying digital information over a real-world view. Autism Spectrum Conditions lend themselves to AR as they can often lead to mental well-being challenges such as stress and anxiety as people with autism often fail to recognise basic facial emotions. Machine learning classifiers can use real-time camera data from AR glasses to infer and inform the wearer of the nearby people's emotions [167]. These AR glasses could greatly help children with autism reduce the daily stress they experience although the machine learning classifier must be improved to recognise faces other than those it has been trained on if it is to be used by the wider population.

There are numerous challenges facing the mainstream use of VR as mental well-being treatment, including the lack of training with only 17% of surveyed licensed psychologists trained to use VR and 38%–46% of those not using VR exposure therapy [168]. To improve VR's mainstream success in improving mental well-being more representative samples and high-quality randomised trials are required to ensure results generalise well in new settings and more psychologists should be trained to use VR exposure therapy.

Virtual reality is now affordable with the tools and technologies required already developed yet its potential to educate people on different coping skills to use in stressful situations has not been fully realised. A potentially controversial topic which raises some concerns is that the recent appearance of VR app stores will allow for VR software to be released without being clinically evaluated, similar to the majority of mental health apps that have been released, and this issue should be addressed before VR software to assist mental well-being enters into mainstream use [68].

### 2.4.2 Biofeedback Therapy

Is it possible to teach people using technology how to improve their mental well-being?

One method to improve mental well-being is biofeedback therapy; this involves monitoring a normal automatic bodily function and then training people to acquire voluntary control of that function. Nolan et al. [169] measured HRV in patients with coronary heart disease as cardiac death is more likely in these patients when stressed. The study recruited 46 patients, of whom 23 undertook HRV biofeedback involving training patients in paced breathing in order to improve their HRV and stress management. The study resulted in patients showing reduced symptoms of psychological stress and depression proving the positive effect of biofeedback training and controlled breathing. Further work is required to investigate whether these findings could be generalised under free-living conditions in community studies.

Another study [170] used biofeedback for general stress management; this biofeedback used a game to encourage users to improve their heart rate and cerebral blood flow control. This study used stress focused questionnaires, a stress marker and a voxel-based morphometric analysis to determine stress allowing the study to conclude that the biofeedback helped reduce daily stress due to the increase in regional grey matter. HRV biofeedback has also been used during the postpartum period after the birth of a child. This study [171] showed that biofeedback helped improve HRV and improve sleep over the 1 month period it was used by 25 mothers. However, the lack of a control group means the study does not definitively show the improvements were due to the biofeedback training alone.

Biofeedback has been shown to have a significant impact in reducing stress during trials, although its effectiveness in real-world stressful situations has not been proven [172]. The possibility of pairing biofeedback training with VR would allow users to practice the techniques learned through biofeedback to reduce stress in a setting they find stressful which would demonstrate the effectiveness of biofeedback. Furthermore, biofeedback requires people to have an understanding, willingness and time to train their body to acquire voluntary control which many people do not possess. Tangible interfaces may solve many of these problems by using sensors to analyse mental state similar to biofeedback, and additionally provide feedback to improve mental well-being in real-time.

### 2.4.3 Real-time Tangible Feedback Interfaces

Can a combination of sensing and feedback technologies be used to improve mental well-being in real-time?

An area of application still in its infancy is technologies that go beyond sensing to additionally provide feedback helping to improve mental well-being. Devices that sense and provide feedback ranging from tangible interfaces to robotics have the possibility to positively impact the broader population who may temporarily experience mental well-being challenges but do not seek professional help. Researchers have developed tangible devices that actively aim to improve mental well-being, these are often paired with sensors and real-world feedback [173] to be automatically provided when required.

A variety of tangible mental well-being devices have been produced by Vaucelle, Bonanni, and Ishii [174] these include: *touch me* which contains multiple vibrotactile motors to provide the sensation of touch; *squeeze me* consisting of a vest to simulate therapeutic holding; *hurt me* consisting of a wearable device that applies a moderated painful stimuli to ground people's senses and *cool me down* a device that heats up to ground people's senses. From the devices developed clinicians believed *hurt me* had the most potential as it could allow for the patient and therapist to better relate to one another, by having the therapist working with the class of pain the patient is experiencing psychologically and externalising viscerally. All of these interfaces have specific purposes such as *hurt me* which may be beneficial for people considering self-harming but not for people suffering from other mental health challenges. A more general mental well-being device is required for people who may experience temporary mental well-being challenges.

It is possible to help improve general mental well-being using small devices with real-time intervention; one such device is Squeeze, Rock and Roll [175]. This device allowed users to simulate rolling behaviours as many people do with a pen when stressed but the device gradually guides the user to reduce their movements and their stress through dynamic tactile feedback. However, while people acknowledged the device helped them relax no stress reduction was found possibly because the device offered very little feedback. Guiding users' behaviours is a novel approach to improve mental well-being although possibly less effective as some people may find the action of rolling or twisting objects relaxing by providing a distraction which can result in mood improvements [176] and is often used as a coping strategy for people suffering from mental health conditions [177].

Haptic feedback is a method of providing feedback that recreates the sense of touch through the use of motors and vibrations; this allows people to experience real sensations which can significantly affect emotional well-being and has been shown to successfully improve mental well-being [178], [179]. Good vibes [180] used a haptic sleeve to provide varying feedback dependent on heart rate readings. A stress test was conducted while the sleeve used dynamic vibrations to help reduce the heart rates of the participants by 4.34% and 8.31% in the two tests compared to the control group. Doppel [181] also used haptic feedback in a wearable device that aimed to reduce stress before public speaking measuring users' heart rates and skin conductance to determine stress. The speed of the vibration was controlled by the user's heart rate providing personalised real-time feedback. When users were told they were to present a speech the skin conductance data showed those wearing the Doppel remained less stressed than the control group. This research shows that haptic feedback can have a substantial positive impact in improving mental well-being and is more successful than guiding user interactions. The advantage of personalised haptic feedback is clear, but more research needs to be conducted to establish the best rate of feedback for individual users.

An alternative to haptic feedback uses deep breathing to improve mental well-being. BioFidget [182] is a self-contained device that uses a heart rate monitor to detect HRV and allows users to train their breathing by blowing on the fidget spinner to reduce stress. Twenty participants stated BioFidget helped them feel relaxed and overall it helped the majority of users improve their HRV showing they were less stressed.

A headband has also been developed that uses EEG combined with machine learning to assess stress by analysing alpha and beta waves as alpha waves decrease when stressed [183] and then uses two low powered massage motors to reduce stress using massage therapy to provide

"*significant reductions in physiological stress*" [184]. The massage motors were tested on 4 participants with 3 of these responding well to the feedback and becoming less stressed showing the possibility for massage therapy to be further utilised in stress reduction devices. However, as the device was only used by 4 participants with a 75% success rate, much more research will need to be conducted to prove it can be used as effectively as haptic feedback.

A different approach to provide real-time feedback is to alert the user regarding their current mental state allowing them to take appropriate measures such as reducing workload or taking time to relax. MoodWings [185] aimed to reduce stress through wing actuations informing users of their current stress levels. Participants wore the device on their arm while ECG and Electrodermal activity (EDA) readings were taken to determine stress. A simulated driving experience was undertaken by participants and once stress was detected the wing movement was manually activated. The results show that MoodWings improve the participants' awareness of their stress, but their awareness further increased their stress as shown by EDA data resulting in the device having a negative effect on users' mental well-being due to its alerting nature. Overall this study demonstrated that sharing data with users needs to be carefully considered [185].

Table 3 summarises the different feedback devices that aim to both detect and help improve mental well-being. Some devices reviewed require manual feedback activation and are not portable, thus making their practical use challenging in real-world settings.

TABLE 3
SUMMARY OF TANGIBLE FEEDBACK DEVICES

| Device | Signal modalities | Features | Validation |
|---|---|---|---|
| Squeeze rock and roll | Force, movement | Dynamic tactile feedback | Minimal stress reduction |
| MoodWings | EKG, EDA, GSM | Moving wings | Resulted in increased stress |
| Good vibes | HR | Vibrotactile feedback | Reduced stress by 4.34% and 8.31% |
| Doppel | HRV, skin conductance | Vibrotactile feedback | 52 users showed lower average skin conductance and state anxiety |
| BioFidget | HRV | Deep breathing | 20/32 stated it helped relaxation, little sensor data |
| Headband | EEG | Massage motors | 3/4 became less stressed |

Communicating with others has a positive mental impact leading to research that remotely connects people through biofeedback. Shared breathing experiences through Breeze using tactile, visual and audio feedback helped to increase the feeling of belonging between connected participants [186]. EmoEcho [187] similarly allowed users to share motion, touch and pulse through haptic feedback with trusted partners to create a remote tangible connection with the aim of improving mental well-being. Stress levels have also been inferred through personal encounters measured using Bluetooth, although reportedly not as accurately as when using physiological sensors [188]. Communication with others is vital to ensure positive mental well-being and while feedback devices that remotely connect individuals appear to improve mental well-being they have only been tested in limited trials.

A novel approach to provide feedback is through the use of robotics such as therapy animals which are most commonly used to reduce loneliness. One example of a robot used for therapy is Paro; a robotic seal that was designed as an easy to use robotic animal that encourages user interaction with its large eyes and soft fur [189]. Tactile sensors allow Paro to understand the location and force of users' touch allowing for the response's magnitude to be relevant to the input. Studies show Paro provided extremely effective therapy as it helped reduce stress in a day service centre for elderly adults [190], increased users' social interactions and improved their reactions to stress in a care home [189]. Paro has been shown to have a great impact in helping reduce stress in elderly adults even with its limited sensors and responses and has the potential to have a wider positive impact on people's mental well-being.

Although most therapeutic robots such as Paro target the elderly, a robotic teddy aimed at reducing stress in young children hospitals has been developed [191]. Rather than relying upon tactile interaction like Paro, this teddy uses vocal interactions which children preferred. The children who used the robotic teddy spent more time playing with it than the comparative virtual or traditional plush teddy, they also had more meaningful interactions and their behaviours conveyed they were emotionally attached to the bear and not stressed. Robotic interactions can have a positive impact on emotional experiences and help reduce stress in both the young and the elderly. Robotic animals could be easily adapted to incorporate additional sensors to automatically detect mental well-being in real-time allowing for more personalised responses to be produced.

Overall, a variety of technologies that both sense mental well-being and provide real-time feedback have been developed. The feedback incorporated in a device requires careful consideration and evaluation to ensure it is effective in improving mental well-being with machine learning being utilised to accurately determine when feedback should be provided.

## 3 REFLECTION AND CHALLENGES OF MENTAL HEALTH TECHNOLOGIES

### 3.1 Discussion of Existing Research

A number of systems to support mental well-being using apps, sensors, tangible interfaces, robotics and biofeedback have been reviewed. A large number of mental well-being apps already exist providing a range of features and functionality with many existing apps aiming to improve traditional self-reporting tools and experience sampling. Apps designed to elicit PROMs provide additional convenience over traditional methods as they can be used anywhere discreetly but self-reporting is subjective and people may fail to report [6] or be less truthful [25] when recording their mental state, showing the benefits of appropriately using objective measurements from sensors even if they are more obtrusive. Recent developments in mHealth apps utilise sensors within smartphones and wearable devices to measure physiological activity allowing mental well-being to be automatically inferred. Currently, this is limited due to the small number of sensors incorporated into such devices but presents a much larger opportunity for continuous mobile mental well-being monitoring [192], [193]. Mobile apps reaffirm the increasing popularity of people wishing to monitor and improve their mental well-being using technological alternatives to traditional techniques. However, currently most mental well-being apps published in the Google Play store and Apple app store have not been evaluated possibly resulting in these apps having unforeseen consequences.

Sensing devices are also increasing in popularity with advancements in physiological and environmental sensors resulting in cheaper and smaller devices promoting extensive use. A range of psychological sensors have been used to detect mental well-being including pulse, HRV, GSR and skin temperature, pairing these with environmental sensors including accelerometer, gyroscope and magnetometer for motion and force sensitive resistors to detect touch enables a wide range of data to be collected to train machine learning models. The ability to pair machine learning algorithms with sensors presents an enormous opportunity allowing for mental well-being to be detected with accuracies exceeding 90% [82], [108]. Empowering sensors with machine learning in a portable interface enables well-being to be continuously monitored without the need to continuously self-report as deep learning models are able to infer mental well-being from the raw data collected. While artificial intelligence has enormous potential in classifying mental states, it does present its own set of challenges as a large amount of labelled data is required to train the model accurately. Furthermore, machine learning models can struggle with predicting future outcomes related to mental illness [194].

Feedback devices aim to advance upon sensing devices by actively improving mental well-being in real-time using varying feedback mechanisms including haptic, visual and auditory [195]. Haptic feedback has been used in multiple devices and often resulted in improved mental well-being especially when the feedback was personalised. Other feedback interfaces aimed to reduce stress using existing techniques such as deep breathing [182], [186], or massage therapy [196]. All these techniques proved to be beneficial in improving mental well-being, demonstrating the need for more widespread adoption of such devices. While some feedback devices incorporated sensors to monitor the impact the feedback had, very little research has been conducted pairing physiological sensors, feedback mechanisms and machine learning into devices that aim to both sense and improve mental well-being in real-time. The effectiveness of the tangible interfaces reviewed drastically varied in mostly small-scale trials, or in some cases no current evaluation showing more evaluation especially real-world trials are required.

## 3.2 Challenges

Applying therapies and translating them into digital or mobile versions is not straightforward as there are many challenges associated with mental well-being technologies.

Privacy is a significant issue as the majority of users want to keep their mental health information private [72]. Users are more cautious regarding sharing their health data making integrating the data with established e-health systems challenging [197]. Ideally data processing should be completed locally although on-device inference is only currently feasible for very limited applications [198]. Furthermore, care needs to be exercised regarding users' privacy with the data collected; ethical guidelines should be abided by, and users should be made aware of the data being collected and how it is being processed.

Given the stigma associated with mental illness, security has to be a high priority for anyone thinking of developing or using mental well-being tools. Concerns about how apps respect privacy and use patient data remain rife, with many mental well-being apps still lacking even basic privacy policies or covertly selling users' mental health information to data brokers. Efforts such as the General Data Protection Regulation (GDPR) in the EU and EEA have attempted to give control to citizens over their personal data by ensuring they are able to access their data and understand how it is being processed [199]. Additionally, the EU Medical Device Regulation (MDR) [200] will require all digital health technologies to pass a conformity assessment and meet safety and performance requirements by 2020.

An issue with some of the discussed devices is users' digital competence as elderly adults generally lack a high level of digital skills which may be required to operate these devices. One study [201] found elderly users preferred wearable devices over mobile phones to report emotions. However, Emoball [72] was a self-contained device rather than a wearable and there was no evidence of digital competence affecting user interactions showing devices to aid mental well-being can be widely adopted.

User adherence and engagement is another crucial problem for well-being devices as users may not immediately see the benefits of such solutions, preventing continued use. Making the devices as small and portable as possible should encourage engagement as it allows them to be used anywhere [74]. The design of the devices must also be carefully considered for widespread use as it must be aesthetically pleasing to ensure the promotion of continuous engagement [202]. However, there should also be considerable debate around how much engagement is necessary to best serve users' particular needs.

Recruiting and incentifying users to test and provide feedback on the use of such devices can be challenging, particularly regarding users' willingness to trial new technologies when it might impact their mental well-being. Users will be required to trial devices to ensure their effectiveness but also to collect data enabling machine learning models to be trained.

An issue with much of the existing research is the lack of control groups and small sample sizes when trialling well-being technologies. Most studies are limited to fewer than 15 participants thus not containing sufficient statistical power to confirm their effectiveness. Furthermore, very few trials collect or test using real-world data as people becoming artificially stressed in trials may not exhibit the same patterns when stressed or suffer from other mental well-being challenges in real-world situations.

Mental well-being can vary widely depending on people's characteristics, and hence it is essential to have a sufficiently representative population sample. On the diagnostic side, one of the biggest issues is mental state sensing: this is inherently subjective and it may be difficult to infer through sensor data alone [203]. Machine learning models could be trained on an individual basis to allow for subjectivity to be taken into account, but this would initially require a vast amount of time and data to be collected from each user before the device could accurately infer well-being which may not be possible if an off-the-shelf device is to be developed. Furthermore, the ability to provide personalised feedback may also require the model to be trained on an individual basis to ensure the most effective feedback for each user is provided. However, as deep learning models require thousands of samples to be sufficiently trained it is difficult to develop a robust deep learning approach for the classification of mental well-being without first developing more accessible data collection tools.

Furthermore, traditional machine learning and feature engineering algorithms may not be sufficiently efficient enough to extract the complex and non-linear patterns generally observed in time series datasets such as those from sensory data. Deep learning can help resolve this issue as the use of a CNN and RNN combined has shown that features can be extracted and classified automatically with LSTM being fundamental in distinguishing time series data.

Sensing mental well-being not only requires accurate machine learning models but also accurate sensors as if the data recorded from the sensors is not reliable the classification from the machine learning model will not be accurate. However, when machine learning classifiers were paired with off the shelf sensors, stress was detected with similar accuracy to clinical grade sensors that are expensive and custom-made [204].

Assuming patients are willing to use instruments used in the domain of assessing mental well-being, the underlying issue of battery life still needs to be addressed. Often IoT devices need to remain small and contain the necessary microcontroller and sensors leaving little room for the battery meaning it will need to be recharged regularly. A possible solution to this would be to only enable specific sensors after other actions have been performed, this means high powered sensors will not have to be continually powered but an additional step is required to collect data. Until batteries with considerably longer battery life are developed, it will remain impractical to continually collect vast amounts of behavioural data. Instead, pragmatic solutions to optimise power consumption are necessary.

If tangible devices are to improve mental well-being, then they must also contain the relevant feedback. There are many challenges to overcome when using sensors and feedback actuators in tangible interfaces to improve mental well-being. One issue is the size of the device as it must contain sensors, a battery and feedback mechanisms such as vibration motors for haptic feedback which make the device large. There are new approaches to provide feedback including Visio-Tactile feedback, that moves liquid metal drops in real-time between electrodes allowing for the feedback to be dynamic and smaller [205]. However, this is very early in development and it may not yet be possible to incorporate it into wearable devices.

Another general challenge is the business opportunity, it will be critical to develop business models based on responsible impact and socially-driven outcomes. There is the possibility of national health systems funding such devices to ease the increasing pressure mental well-being challenges have on health care, but a lack of government funding may prevent this.

Overall there are many challenges to overcome when developing tangible mental well-being devices ranging from privacy issues to technological problems, but new regulations along with technological advancements should help reduce the difficulties these challenges impose.

### 3.3 Opportunities

#### 3.3.1 User Feedback

The opportunities new technologies present to monitor and improve mental well-being were explored during focus groups at a school for students with severe, profound and complex learning and physical disabilities in Nottingham, UK. Mobile well-being apps were discussed although not used by the participants due to their complexity as many participants had fine and gross motor control issues making touchscreens challenging to use, demonstrating the need to develop tools to target specific sub-categories. Alternatives to mobile apps such as tangible interfaces and virtual reality show more potential for this user group as they are easier to handle and operate.

Existing examples of mental well-being tangible interfaces were discussed to explore the opportunities they present. Participants liked the portability of tangible devices and the different methods of interactions compared with smartphones. Participants were excited by the concept of devices being able to infer their mental well-being as many had trouble recording their emotions. The possibility for devices to improve mental well-being was also intriguing as the participants had not used such devices, demonstrating the requirement for tangible interfaces to sense and improve mental well-being.

Wearable devices were considered to be useful as they remove any requirement for fine motor control. Different motor control levels were examined in a separate group which showed some participants' inability to tightly grip objects while others had difficulty relaxing their muscles. This demonstrates it may not be possible to develop a single tangible device aimed at all people suffering mental well-being challenges; separate interfaces may need to be developed targeting different groups of people.

Cost was a key factor discussed during the focus group as the school and individuals would need the device to be inexpensive if it was to become adopted into practice. Durability was another issue raised as devices can often be used in unintended ways which must be considered during design and development. This focus group demonstrates the need for a range of technological solutions to address mental well-being issues as a one-size-fits-all solution could not feasibly address all mental well-being issues for all potential users, especially those whose mental well-being issues are often diagnostically overshadowed. The session concluded that for mental well-being, tangible interfaces demonstrate the most potential to both express feelings as well as actively improve mental well-being but cost, durability and ergonomics need to be prioritised.

*3.3.2 Advancements to Enable Real-time Intervention*

Recently there have been many developments in the tools required to develop devices to sense and improve mental well-being in real-time including the required microprocessors and sensors. Numerous System on Chip (SoC) devices are now available that are capable of reading data from sensors as well as processing data in (near) real-time. Microcontrollers such as the Arduino platform are currently limited in terms of computational power towards complex data processing; however, the popularity of mobile phones enables microcontrollers to export the data to be processed externally.

Additionally, advances in mobile phones and edge computing allow for machine learning to classify the data collected from sensors locally. Many machine learning frameworks have been developed to run on low powered devices including TensorFlow lite which displayed high performance in both single inference latency and CPU-optimized continuous throughput when tested on Android phones [190]. It is now possible to run TensorFlow models on smartphones and devices such as the Raspberry Pi, enabling interfaces powered by these devices to use deep learning to infer mental well-being in real-time. Recently, a personalised transfer learning approach to infer stress was performed locally using a Raspberry Pi achieving up to 93.9% accuracy [206]. These advancements allow for small, portable, unobtrusive devices to be developed which can utilise deep learning to improve people's mental well-being in real-time while preserving privacy.

## 4 CONCLUSION

Different methods to sense and improve mental well-being have been considered including apps, sensing devices, behaviour changing tools and real-time intervention devices. Tangible interfaces present a substantial opportunity for mental well-being devices as they have the capability to both sense mental well-being and provide interventional feedback. Sensors to detect well-being can now be incorporated into small devices and advances in deep learning allow for the raw data to be classified accurately on-device allowing for real-time personalised feedback.

Personalising the feedback, tangible interfaces can provide presents a great opportunity towards delivering precision medicine and offering patient-specific suggestions and interventions, a premise which has so far not been delivered at scale in healthcare decision support applications. Personalised feedback also removes the assumption many existing tangible interfaces have made by creating a one-size-fits-all device as different people suffering from poor mental well-being may prefer and respond better to different interventions.

There are numerous challenges associated with mental well-being technologies such as the size of the device, data collection, privacy, and battery life; however, recent technological advances have truly revolutionized the way forward for small devices to monitor and improve mental well-being. Wearable devices would enable easier collection of physiological data. However, ensuring all of the electronics are sufficiently small to be contained within a wrist-worn device may reduce battery life and increase costs.

Tangible user interfaces go beyond the capabilities that mobile apps can offer but have not yet been fully explored. There is relatively little research conducted in the use of tangible devices to both infer and improve mental wellbeing in real-time. Many existing studies rely on small sample trials conducted over a short period of time and without a suitable control condition, making it challenging to evaluate their long-term effectiveness. More rigorous studies need to be conducted to provide robust evidence for the alleged capabilities tangible interfaces possess to enable such technology to be modified, scaled and culturally adapted to serve the global population.

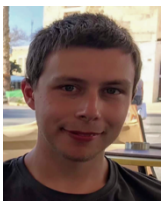

**Kieran Woodward** graduated from Nottingham Trent University (NTU) with a First Class BSc (Hons) degree in Information and Communications Technology (2016) and MSc Computing Systems (2017). He is currently pursuing his PhD at NTU researching the use of tangible user interfaces and on-device machine learning to infer mental well-being in real-time.

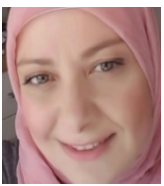

**Eiman Kanjo** is an Associate Professor in Mobile Sensing & Pervasive Computing at Nottingham Trent University. She is a technologist, developer and an active researcher in the area of mobile sensing, smart cities, spatial analysis, and data analytics, who worked previously at the University of Cambridge, Mixed Reality Laboratory, University of Nottingham and the International Centre for Computer Games and Virtual Entertainment, Dundee. She authored some of the earliest papers in the area of Mobile Sensing and currently carries out work in the area of Digital Phenotyping Smart cities, Mental Health and the Internet of Things for Behaviour Change in collaboration with many industrial partners and end-user organizations.

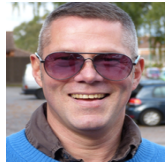

**David J Brown** is Professor of Interactive Systems for Social Inclusion at Nottingham Trent University. He is Principal Investigator two EU H2020 Grants (MaTHiSiS and No One Left Behind) to investigate the use of sensor data to understand the emotional state of learners to provide personalised learning experiences, and how game making can enhance the engagement of students with learning disabilities and autism. He is also Co-Investigator to EPSRC Internet of Soft Things, to investigate the impact of networked smart textile objects on young people's wellbeing.

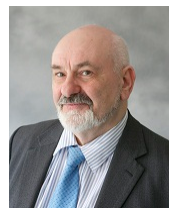

**T. Martin McGinnity** (SMIEEE, FIET) currently holds a part-time Professorship in both the Department of Computing and Technology at Nottingham Trent University (NTU), UK and the School of Computing, Engineering and Intelligent Systems at Ulster University, N. Ireland. He was formerly Pro Vice Chancellor, Head of the College and Dean of the School of Science and Technology at NTU, Head of the School of Computing and Intelligent Systems and Director of the Intelligent Systems Research Centre in Ulster University. He is the author or coauthor of over 330 research papers and leads the Computational Neuroscience and Cognitive Robotics research group at NTU.

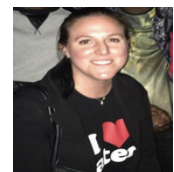

**Becky Inkster** is a neuroscientist, passionate about everything from cells to phones, genes to jewellery, hip-hop to hippocampi. Becky is affiliated with the University of Cambridge, Creator of the Digital Innovation in Mental Health conference, Co-founder of Hip Hop Psych, and holds several advisory positions (e.g., The Alan Turing Institute; The Lancet Digital Health; Mental Health and Suicide Prevention, Global Advisor, Facebook; Clinical Advisory Board Member, TalkLife; Advisor, Wysa; AI Global Governance Commission; IBM Watson AI XPrize)

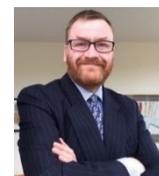

**Donald MacIntyre** graduated from the University of Edinburgh with a BSc (Hons) Medical Science (1994), MBChB (1996) and MD (2013). He is a consultant general psychiatrist in NHS Lothian and is seconded to NHS 24 as Associate Medical Director (Mental Health). He was appointed as an NHS Research Scotland Fellow in 2014, Honorary Reader in Psychiatry at the University of Edinburgh in 2017, and made a Fellow of the Faculty of Clinical Informatics in 2019. He is interested in implementing technology enabled mental health care.

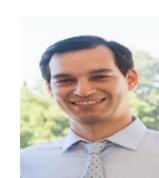

**Athanasios Tsanas ('Thanasis'), SMIEEE,** completed a BSc in Biomedical Engineering at the Technological Educational Institute of Athens (2005), a BEng in Electrical Engineering and Electronics at the University of Liverpool (2007), an MSc in Signal Processing and Communications at the Newcastle University (2008), and a DPhil (PhD) in Applied Mathematics, Mathematical Institute, University of Oxford (2012). He was previously a post-doctoral research fellow at the University of Oxford (2012-2016). He is an Assistant/Associate Prof. in Data Science (tenure-track) at the Medical School, University of Edinburgh, a Lecturer in Statistical Research Methods at the University of Oxford, and a Fellow of the Royal Society of Medicine.